\documentclass[final]{elsart}

\usepackage[dvips]{color}
\usepackage{epsfig}
\usepackage{amsmath}
\usepackage{qsymbols}
\usepackage{natbib}

\begin{document}

\begin{frontmatter}

\title{Stress Field at a Sliding Frictional Contact: Experiments and Calculations}
\author{J. Scheibert\corauthref{cor1}}\footnote{Now in Physics of Geophysical Processes, University of Oslo, Oslo, Norway},
\ead{julien.scheibert@fys.uio.no}
\author{A. Prevost},
\author{G. Debr\'egeas},
\author{E. Katzav}\footnote{Now in Department of Mathematics, King's College London, United Kingdom},
\author{M. Adda-Bedia}

\corauth[cor1]{Corresponding author}
\address{Laboratoire de Physique Statistique de l'ENS, UMR 8550, CNRS/ENS/Universit\'e Paris 6/Universit\'e Paris 7, 24 rue Lhomond, 75231 Paris, France}

\begin{abstract}
A MEMS-based sensing device is used to measure the normal and tangential stress fields at the base of a rough elastomer film in contact with a smooth glass cylinder in steady sliding. This geometry allows for a direct comparison between the stress profiles measured along the sliding direction and the predictions of an original \textit{exact} bidimensional model of friction. The latter assumes Amontons' friction law, which implies that in steady sliding the interfacial tangential stress is equal to the normal stress times a pressure-independent dynamic friction coefficient $\mu_d$, but makes no further assumption on the normal stress field. Discrepancy between the measured and calculated profiles is less than 14\,$\%$ over the range of loads explored. Comparison with a test model, based on the classical assumption that the normal stress field is unchanged upon tangential loading, shows that the exact model better reproduces the experimental profiles at high loads. However, significant deviations remain that are not accounted for by either calculations. In that regard, the relevance of two other assumptions made in the calculations, namely (i) the smoothness of the interface and (ii) the pressure-independence of $\mu_d$ is briefly discussed.
\end{abstract}

\begin{keyword}
Contact mechanics \sep Layered rubber material \sep Friction \sep MEMS \sep Integral transforms
\PACS 46.55.+d \sep 81.40.Pq \sep 85.85.j
\end{keyword}

\end{frontmatter}

\section{Introduction}

The sliding contact between non-conforming elastic bodies is a classical problem in contact mechanics (\cite{Cattaneo-RANL-1938,Mindlin-JApplMech-1949,Johnson-CUP-1985,Hills-Nowell-Kluwer-1994}). Knowledge of the surface and subsurface stress field in such systems is central to solid friction, seismology, biomechanics or mechanical engineering. Typical applications include hard disk drives (e.g. \cite{Talke-Wear-1995}), tribological coatings (e.g. \cite{Holmberg-Matthews-Ronkainen-Tribolint-1998}), train wheels on rails (e.g. \cite{Guagliano-Pau-TribolInt-2007}), human joints (e.g. \cite{Barbour-Barton-Fisher-JMaterSci-1997}) and tactile perception (e.g. \cite{Howe-Cutkosky-IEEETransRobAutom-1993, Scheibert-Leurent-Prevost-Debregeas-Science-2009}).

Theoretically, calculations of the contact stress field in the quasi-static steady sliding regime have been performed for both homogeneous (\cite{Poritsky-JApplMech-1950,Bufler-IngArch-1959,Hamilton-Goodman-JApplMech-1966,Hamilton-JMechEngSci-1983}) and layered elastic half-spaces (\cite{King-OSullivan-IntJSolidsStructures-1987,Nowell-Hills-IntJSolidsStruct-1988,Shi-Ramalingam-SurfCoatTech-2001}), for cylindrical (\cite{Poritsky-JApplMech-1950,Bufler-IngArch-1959,Hamilton-Goodman-JApplMech-1966,King-OSullivan-IntJSolidsStructures-1987,Nowell-Hills-IntJSolidsStruct-1988}), circular (\cite{Hamilton-Goodman-JApplMech-1966,Hamilton-JMechEngSci-1983}) or elliptical (\cite{Shi-Ramalingam-SurfCoatTech-2001}) contacts. These calculations assume a locally valid Amontons' friction law, stating that everywhere within the sliding contact region, the interfacial tangential stress $q=\mu_d p$ with $p$ being the interfacial normal stress and $\mu_d$ the dynamic friction coefficient. Up to now, no quantitative comparison between such calculations and experimental stress fields has been performed. The present work first aims at at filling this lack, by taking advantage of a recently proposed experimental method (\cite{Scheibert-Edilivre-2008,Scheibert-Prevost-Frelat-Rey-Debregeas-EPL-2008,Scheibert-Leurent-Prevost-Debregeas-Science-2009}), which allows for direct measurements of the stress field at the rigid base of a frictional elastomer film.

For such a layered system, no exact stress calculation in a steady sliding contact has been provided up to now neither. All previous works indeed rely on the classical Goodman's assumption which states that the normal displacements at the interface due to tangential stress are negligible (\cite{Goodman-JApplMech-1962}). This implies in particular that the interfacial pressure field is unaltered when a macroscopic tangential load is applied.
For a contact between elastic half-spaces, such a normal/tangential decoupling occurs only if (i) both materials are identical, (ii) both are incompressible or (iii) one of both is perfectly rigid while the other is incompressible (\cite{Bufler-IngArch-1959,Dundurs-Bogy-JApplMech-1969}). For layered systems, Goodman's assumption is never strictly true. However, it is expected to be increasingly valid (i) the higher the Poisson's ratio (\cite{Kuznetsov-Wear-1978}), (ii) the lower the ratio of the contact size $a$ over the film thickness $h$ or (iii) the lower the friction coefficient. Rigorously, one has to keep in mind that Goodman's assumption does not have any physical ground since it does not impose the continuity of the normal displacements between the two solids in contact. The present work presents an exact stress analysis which, for a single linear elastic incompressible layer (film) under plane strain conditions, goes beyond the classical description by relaxing Goodman's assumption.

In section \ref{Set-up}, we describe the experimental setup along with the calibration of the apparatus. In Section \ref{Measurements}, we present both the normal and tangential stress profile measurements at the base of the elastomer film obtained with a cylinder-on-plane contact in steady sliding. In Section \ref{model}, we present the exact model for the quasi-static steady sliding of a rigid circular frictional indentor against the film. In Section \ref{discussion}, the results of this exact calculation are compared to that of a semi-analytical test model implemented with Goodman's assumption. The measurements are directly compared to both models and discussed.

\section{Set-up and Calibration}\label{Set-up}

\begin{figure}[ht!]
\includegraphics[width=\columnwidth]{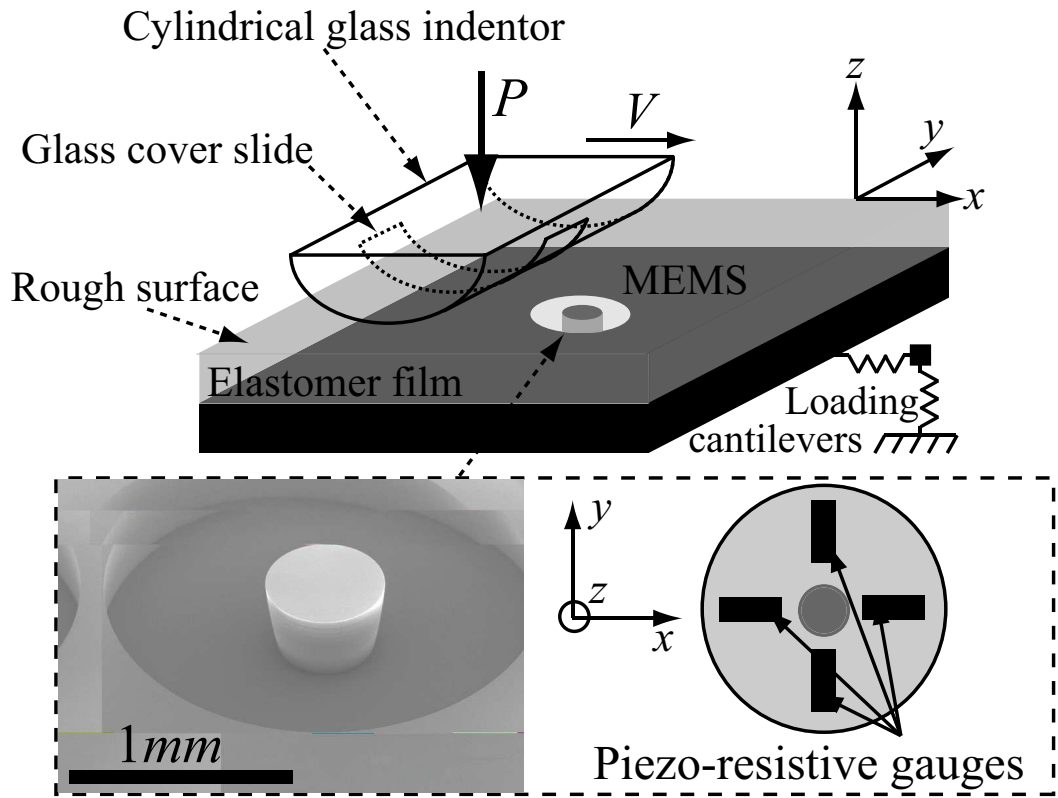}
\caption{Sketch of the experimental setup. A cylindrical glass lens (radius of curvature 129.2\,mm) to which is glued a glass cover slide is driven along the $x$ direction against a rough, nominally flat PDMS elastomer film (uniform thickness $h$\,=\,2\,mm, lateral dimensions 50$\times$50\,mm) at a constant prescribed normal load $P$ and a constant velocity $V$ using a linear DC servo-motor (LTA-HS, Newport). The local normal and tangential stress at the rigid base of the film, respectively $\sigma_{zz}$ and $\sigma_{xz}$, are measured by a MEMS force sensor, whose sensitive part is shown in the lower inset (left hand), along with a sketch (right hand) showing the piezo-resistive gauges implementation within the Silicon membrane. $P$ and the tangential load $Q$ applied on the film are measured through the extension of two orthogonal loading cantilevers (normal stiffness 641$\pm$5\,N.m$^{-1}$, tangential stiffness 51100$\pm$700\,N.m$^{-1}$) by capacitive position sensors (respectively MCC30 and MCC5, Fogale nanotech).\label{schemamanip}}
\end{figure}

Local contact stress measurements are performed with a Micro Electro Mechanical Sytem (MEMS) force sensor embedded at the rigid base of an elastomer film (Fig. \ref{schemamanip}). The MEMS' sensitive part (Fig. \ref{schemamanip}, inset) consists of a rigid cylindrical post (diameter 550\,$\mu$m, length 475\,$\mu$m) attached to a suspended circular Silicon membrane (radius 1\,mm, thickness 100\,$\mu$m, 330\,$\mu$m below the MEMS top surface). When a force is applied to the post, the resulting (small) deformations of the membrane are measured via four couples of piezo-resistive gauges embedded in it and forming a Wheatstone bridge (see inset of Fig. \ref{schemamanip}). The MEMS thus allows to measure simultaneously the applied stress along three orthogonal directions, averaged over the MEMS's millimetric extension, in a way that will be determined through calibration.

In the present experiments, the MEMS sensor is located at the rigid base of a rough, nominally flat elastomer film of uniform thickness $h$\,=\,2\,mm ($``~4$ times larger than the post's diameter) and lateral dimensions 50$\times$50\,mm. The elastomer is a cross-linked Poly(\-Di\-Methyl\-Siloxane) (PDMS, Sylgard 184, Dow Corning) of Young's modulus $E$\,=\,2.2$\pm$0.1\,MPa and Poisson's ratio $\nu$\,=0.5 (\cite{PolyDataHandbook-OUP-1999}). The ratio of its loss over storage moduli, measured in a parallel plate rheometer, remains lower than $``~0.1$ for frequencies smaller than 1\,kHz (\cite{Scheibert-Edilivre-2008}). In this range the PDMS elastomer can thus be considered as purely elastic. The film is obtained by pouring the cross-linker/PDMS liquid mix directly on the sensitive part of the MEMS (cylindrical post and membrane) so that the resulting elastic film is in intimate contact with the MEMS sensitive part. The parallelepipedic mold used in this process is topped with a Poly(\-Methyl\-Meth\-Acrylate) plate roughened by abrasion with an aqueous solution of Silicon Carbide powder (mean diameter of the grains 37\,$\mu$m). After curing at room temperature for at least 48 hours and demolding, the resulting $rms$ surface roughness is measured with an interferential optical profilometer (M3D, Fogale Nanotech) to be 1.82$\pm$0.10\,$\mu$m. This roughness is sufficient to avoid any measurable pull-off force against smooth glass indentors, as discussed in \cite{Fuller-Tabor-ProcRSocLondA-1975}. When the film is put in contact against an indentor, the normal and tangential loads applied, respectively $P$ and $Q$ are measured through the extension of two orthogonal loading cantilevers (normal stiffness 641$\pm$5\,N.m$^{-1}$, tangential stiffness 51100$\pm$700\,N.m$^{-1}$) by capacitive position sensors (respectively MCC30 and MCC5, Fogale nanotech).

The stress sensing device (MEMS with its PDMS film) has been calibrated in an earlier work (\cite{Scheibert-Prevost-Frelat-Rey-Debregeas-EPL-2008}), for the normal stress only. The method is recalled here and extended to the tangential stress. The surface of the film is indented with a rigid cylindrical rod of diameter 500\,$\mu$m, under a normal load $P$. With this flat punch indentor, all sensor outputs are found to be linear with $P$. By successively varying the position of this rod along the $x$ direction, and assuming homogeneity of the surface properties of the film, the radial profiles of the normal and tangential output voltages, respectively $U_{zz}(x)$ and $U_{xz}(x)$, are constructed point by point. These profiles are then compared to the results of finite elements calculations (Software Castem 2007) for the stress $\sigma_{zz}$ and $\sigma_{xz}$ at the base of a smooth axi-symmetrical elastic film (with the same elastic moduli and thickness as in the experiment) perfectly adhering to its rigid base and submitted to a prescribed normal displacement over a central circular area of diameter 500\,$\mu$m. For frictionless conditions, these numerical results could have been obtained semi-analytically by using the model developed in \cite{Fretigny-Chateauminois-JPhysD-2007} but finite elements calculations have been preferred because they allowed for variable boundary conditions. As expected for contact regions of dimensions smaller than the film thickness, the stress calculated at the base of the film are found to be insensitive to the frictional boundary conditions.

The vertical dimensions of the MEMS being smaller than the thickness of the elastomer film, one can ignore the stress field modifications induced by the MEMS 3D structure and consider that the base of the film is a plane. We can then relate the measured output voltage $U$ to the stress field at the base of the film $\sigma$ by writing down that
\begin{equation}
U_{\alpha z}(x,y)=A_{\alpha z} G_{\alpha z} \otimes \sigma_{\alpha z} (x,y) \label{convolalphaz}
\end{equation}
where $\alpha = x$ or $z$. $A_{zz}$ and $A_{xz}$ are conversion constants (units of $mV / Pa$), $G_{zz}$ and $G_{xz}$ are normalized apparatus functions and $\otimes$ is a convolution product. Note that we use the sign convention that $\sigma_{zz}$ is positive for compressive loading. Eqs. \ref{convolalphaz} implicitly assume decoupling between the MEMS outputs. This has been checked to be true for the bare sensor by submitting it to either a uniform pressure or a pure tangential load applied directly on the Silicon cylindrical post. When the MEMS is embedded in the elastomer film, this remains true for the normal output, as checked by applying a uniform pressure at the surface of the film. The analogous check for the tangential output is not possible because any tangential stress applied on the film surface results in tangential stress as well as normal stress gradients at its base, which can not be measured separately since they induce the same deformation mode of the MEMS Silicon membrane. One can still use Eqs. \ref{convolalphaz} in the limit of contact configurations involving small pressure gradients. This is the case when one uses indentors with large radius of curvature such as the cylinders considered in the rest of this study. In this limit, the tangential output is likely to be insensitive to normal stress since the Silicon sensor is much stiffer than the elastomer.

In Fourier space, Eqs. (\ref{convolalphaz}) become
\begin{equation}
A_{\alpha z} G_{\alpha z}(x,y)=\mathcal{F}^{-1}\left(\frac{\mathcal{F}\left\{U_{\alpha z}\right\}(f_{x},f_{y})}{\mathcal{F}\left\{\sigma_{\alpha z}\right\}(f_{x},f_{y})}\right)(x,y) \label{Galphaz}
\end{equation}
where $\mathcal{F}$ is the bidimensional spatial Fourier Transform, $\mathcal{F}^{-1}$ its inverse, and $f_{x}$, $f_{y}$ are the spatial frequencies in the $x$, $y$ directions respectively. The $U_{zz}(x,y)$, $U_{xz}(x,y)$, $\sigma_{zz}(x,y)$ and $\sigma_{xz}(x,y)$ fields are built from the corresponding profiles along the $x$ axis, assuming axi-symmetry, and then transformed using a Fast Fourier Transform (FFT) algorithm. The rapid decay of $\mathcal{F}\left\{\sigma_{zz}\right\}$ and $\mathcal{F}\left\{\sigma_{xz}\right\}$ with increasing spatial frequency yields a divergence of the ratio in Eqs. (\ref{Galphaz}). To circumvent this difficulty, a white noise of amplitude 10 times weaker than the weakest relevant spectral component is added to both terms of the ratio before applying the FFT. The result is found to be insensitive to the particular amplitude of this white noise. $A_{zz}$ and $A_{xz}$ are determined so that the integrals of $G_{zz}$ and $G_{xz}$ are equal to 1. The integrals of both $U_{xz}$ and $\sigma_{xz}$ being zero, $G_{xz}$ is determined up to an additive constant, which was taken such that $G_{xz}$ vanishes far from the MEMS location.

\begin{figure}[ht!]
\includegraphics[width=\columnwidth]{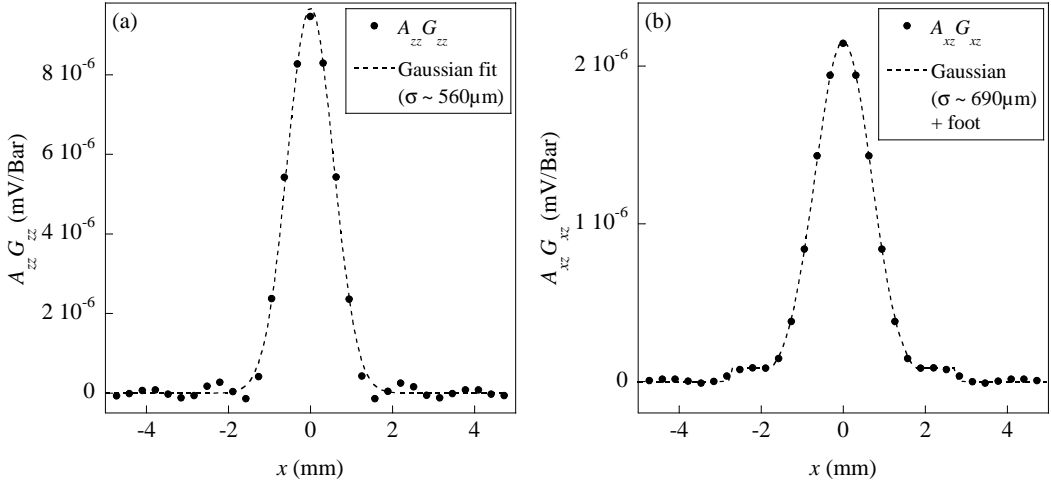}
\caption{Apparatus functions $A_{zz}G_{zz}$ and $A_{xz}G_{xz}$ of the MEMS sensor for (a) the normal stress and (b) the tangential stress, respectively. Black dots are the results of the calibration method. Dashed lines are the approximated apparatus functions used for subsequent calculations.\label{calib}}
\end{figure}

Both $G_{zz}$ and $G_{xz}$ exhibit a bell shape with a typical width of the order of 600\,$\mu$m, comparable to the lateral dimension of the sensitive part of the MEMS (Fig. \ref{calib}). For subsequent calculations, $G_{zz}$ is approximated by a gaussian of standard deviation 561\,$\mu$m (Fig. \ref{calib}(a)). The shape of $G_{xz}$ is more complex and is therefore approximated by a gaussian of standard deviation 688\,$\mu$m decorated by a rectangular foot of lateral extent 2.7\,mm and amplitude 4.1\,\% of the maximum amplitude of $G_{xz}$ (Fig. \ref{calib}(b)). We checked that a simple gaussian approximation of $G_{xz}$was not sufficient to reproduce the measured $U_{xz}$ profile when convoluted with $\sigma_{xz}$.

\begin{figure}[ht!]
\includegraphics[width=\columnwidth]{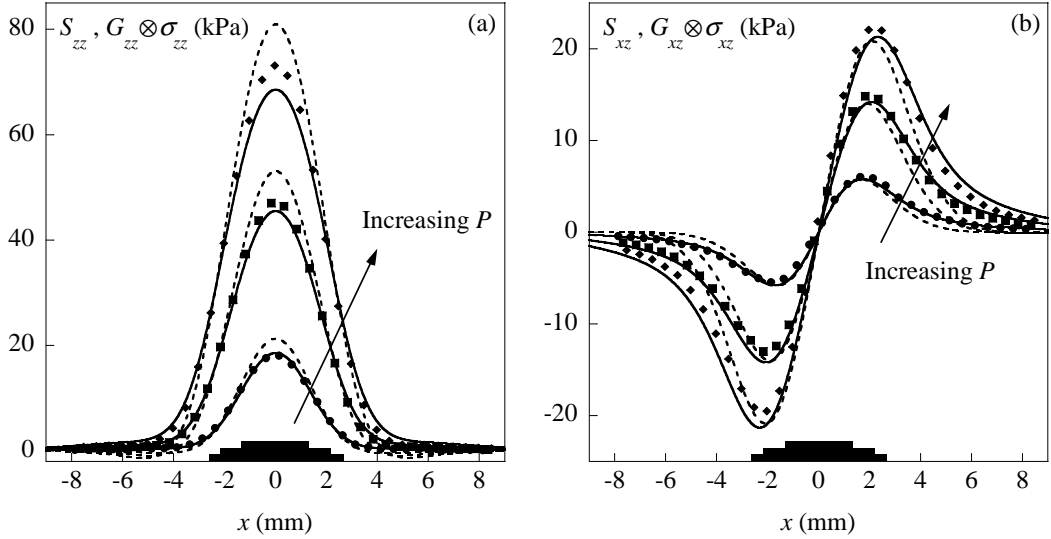}
\caption{Validation of the calibration procedure. Measured stress profiles under normal loading by the rigid cylinder ($P$\,=\,0.69\,N ($`(!)$), 1.72\,N ($`[!]$) and 2.75\,N ($\blacklozenge$) (a) normal stress $S_{zz}(x)$ and (b) tangential stress $S_{xz}(x)$. Comparison is made with $G_{zz} \otimes \sigma_{zz}(x)$ and $G_{xz} \otimes \sigma_{xz}(x)$ for $\mu_s$\,=\,0 (solid lines) and $\mu_s = \infty$ (dashed lines). The black rectangular patches on the $x$ axis represent the contact widths (3.00, 4.50 and 5.34\,mm for $P$\,=\,0.69, 1.72 and 2.75\,N, respectively) obtained from the finite elements calculations for $\mu_s$\,=\,0.\label{static}}
\end{figure}

To validate this calibration procedure, the stress profiles $S_{zz}(x)=U_{zz}(x)/A_{zz}$ and $S_{xz}(x)=U_{xz}(x)/A_{xz}$ in the $x$ direction for cylinder-on-plane contacts under a pure normal load are measured point by point in the same way as for the rod indentation. Note that the sign convention for $S_{zz}$ is the same as for $\sigma_{zz}$. The indentor is a glass cover slide (thickness 150\,$\mu$m, $y$ dimension $L$\,=\,8\,mm) glued with a very thin film of cyanoacrylate onto the cylindrical part of a plano-convex cylindrical glass lens of radius of curvature 129.2\,mm (Fig. \ref{schemamanip}). The contact length in the $y$ direction is therefore 8\,mm, a dimension which is large enough to create locally, at its center, a $y$ invariant stress state, but small enough to make the contact insensitive to flatness imperfections at the scale of the elastomeric film lateral size. Both the glass and the PDMS surfaces are passivated using a vapor-phase silanization procedure which reduces and homogenizes the surface energy (\cite{Chaudhury-Whitesides-Langmuir-1991}). Each contact is formed using the following loading sequence. The indentor is pressed against the PDMS film up to the prescribed load $P$ within 2\,\% relative error. Due to the associated tangential displacement of the extremity of the normal cantilever, a significant tangential load $Q$ is induced. From this position, the contact is renewed by manual separation which results in a much smaller but finite $Q$. To correct for this residual load, the indentor is displaced a few micrometers tangentially down to $Q$\,=\,0. Finite elements calculations using the same geometrical and loading conditions are performed with both zero and infinite static friction coefficients $\mu_s$ in order to provide limiting boundary conditions. The calculated stress profiles $\sigma_{zz}(x)$ and $\sigma_{xz}(x)$ at the base of the elastic film are then convoluted by the apparatus functions $G_{zz}$ and $G_{xz}$ to allow for comparison with the corresponding experimental measurements. The value  $A_{zz}$\,=\,19.00\,mV/Bar obtained by deconvolution allows for the pressure profile measurements to lie between the $\mu_s$\,=\,0 and $\mu_s = \infty$ limiting calculated profiles, in the whole load range further used in this work (Fig. \ref{static}(a)). An equally good agreement is obtained for the tangential stress profiles with $A_{xz}$\,=\,7.95\,mV/Bar, a value 7\,\% higher than the one determined by deconvolution\footnote{This difference is very likely due to the above mentioned fact that the MEMS' tangential output is sensitive to pressure gradients over the size of the sensor. These gradients are estimated to represent less than 6\,\% of the tangential output for the rod indentation situation used to determine $A_{xz}$. For the large cylinder-on-plane contacts under normal loading that are considered in this calibration, the normal stress gradients vanish with increasing normal load. They represent at most 4\,\% of the tangential output over the whole range of $P$ used here.} (Fig. \ref{static}(b)). We checked that $G_{yz}=G_{xz}$ and $A_{yz}=A_{xz}$. These apparatus functions are assumed to remain valid for contacts in the steady sliding regime\footnote{In steady sliding, the normal stress gradients represent a decreasing proportion of the tangential output with increasing normal load, less than 16\,\% for $P$\,=\,0.34\,N, less than 9\,\% for $P$\,=\,0.69\,N, down to less than 4\,\% over $P$\,=\,2.40\,N.}.

\section{Steady sliding measurements}\label{Measurements}

The steady sliding experiments are carried out as follows. Prior to sliding, contacts are prepared under normal load only, ranging from 0.34\,N to 2.75\,N, using the loading sequence described in Section \ref{Set-up}. The cylindrical indentor is then translated tangentially over 20\,mm along the positive $x$ direction at constant velocity $V$ between 0.2\,mm.s$^{-1}$ and 2\,mm.s$^{-1}$. Reproducibility is such that $Q(t)$ differs from less than 1\,\% between two successive experiments (same $P$ and $V$). The signals display a short transient followed by a steady sliding regime for which both $Q(t)$ and $P(t)$ exhibit uncorrelated fluctuations of relative amplitude smaller than 4\,\%. This observation indicates that the surface properties can be considered as homogeneous throughout the explored area. It allows us to derive the stress profiles along the sliding direction directly from the MEMS signals through the relation $S_{\alpha z}(x)=U_{\alpha z}(Vt)/A_{\alpha z}$  (with $\alpha = x$, $y$ and $z$).

\begin{figure}[ht!]
\includegraphics[width=\columnwidth]{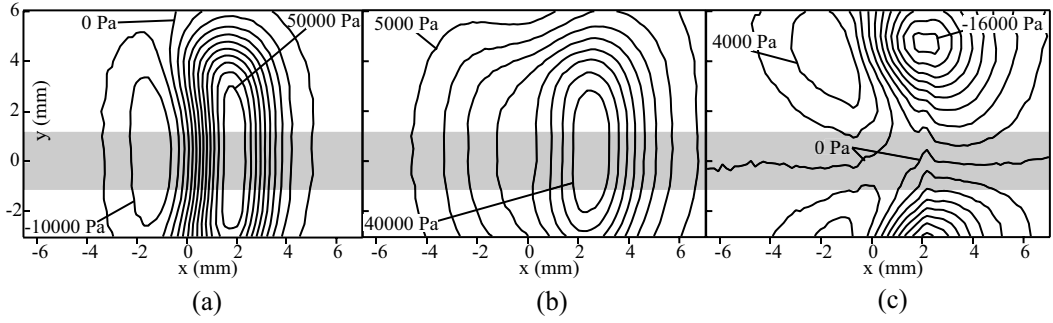}
\caption{Measured stress field for a cylinder-on-plane contact in steady sliding regime at $V$\,=\,0.4\,mm.s$^{-1}$ and $P$\,=\,1.72\,N. (a) Normal stress $S_{zz}$ (b) Tangential stress along the direction of movement $S_{xz}$ (c) Tangential stress orthogonal to the direction of movement $S_{yz}$. Lines are iso-stress curves obtained by interpolation of 19 $x$-profiles made of 10000 data points each. The shaded zone defines the region in which quasi 2-dimensional conditions are met. The measured field is not centered on the contact due to limitations in the movement of the translation stage.\label{CP2D}}
\end{figure}

Figure \ref{CP2D} shows the measured steady sliding stress field for all 3 components $S_{zz}$, $S_{xz}$ and $S_{yz}$ at $P$\,=\,1.72\,N and $V$\,=\,0.4\,mm.s$^{-1}$. They have been constructed from the interpolation of 19 profiles along $x$ at different locations with respect to the MEMS, with 0.5\,mm steps along the $y$ axis. Each profile is made of 10000 data points, one every 2\,$\mu$m. The line $x$\,=\,0 corresponds to the center of the cylinder-on-plane stress profile measured under normal load, while the axis $y$\,=\,0 corresponds to the symmetry line of the steady-state stress field. These fields are to a good approximation $y$ invariant over a width of a few millimeters (shaded region in Fig. \ref{CP2D}) comparable to the extension of the MEMS field of integration. This observation allows us to consider that the $x$ profiles at $y$\,=\,0 provide an experimental realization of a 2-dimensional (i.e. $y$ invariant) cylinder-on-plane friction experiment. In the following we will focus on these profiles and compare them with calculated stress profiles under plane strain conditions. For a given $P$, the profiles obtained with a sliding velocity $V$ in the range 0.2\,$<$\,$V$\,$<$\,2.0\,mm.s$^{-1}$ are almost undistinguishable. Thus, in the following, only the profiles obtained with $V$\,=\,1.0\,mm.s$^{-1}$ are shown.

\begin{figure}[ht!]
\includegraphics[width=\columnwidth]{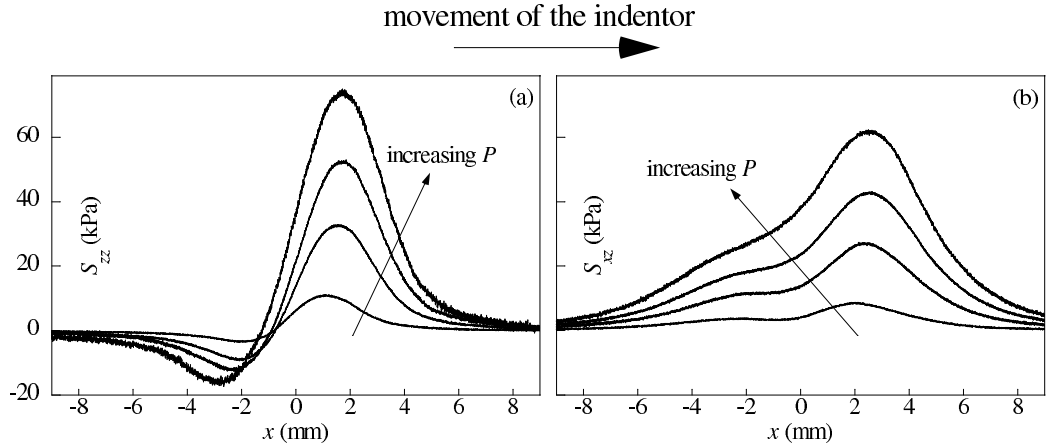}
\caption{Measured stress profiles at $y$\,=\,0 for a cylinder-on-plane contact in steady sliding regime at $V$\,=\,1\,mm.s$^{-1}$ for $P$\,=\,0.34, 1.03, 1.72 and 2.75\,N. (a) Normal stress $S_{zz}$ (b) Tangential stress $S_{xz}$ along the direction of movement.\label{dyn-V1-4P}}
\end{figure}

Figure \ref{dyn-V1-4P} shows the measured stress profiles $S_{zz}(x)$ and $S_{xz}(x)$ for 4 different normal loads. For both components, the profiles exhibit a similar shape with a maximum at the leading edge of the moving indentor whose amplitude increases with $P$. The tangential component is positive throughout the contact whereas the normal component exhibits a negative minimum at the trailing edge.

\section{Exact model}\label{model}

\begin{figure}[ht!]
\includegraphics[width=\columnwidth]{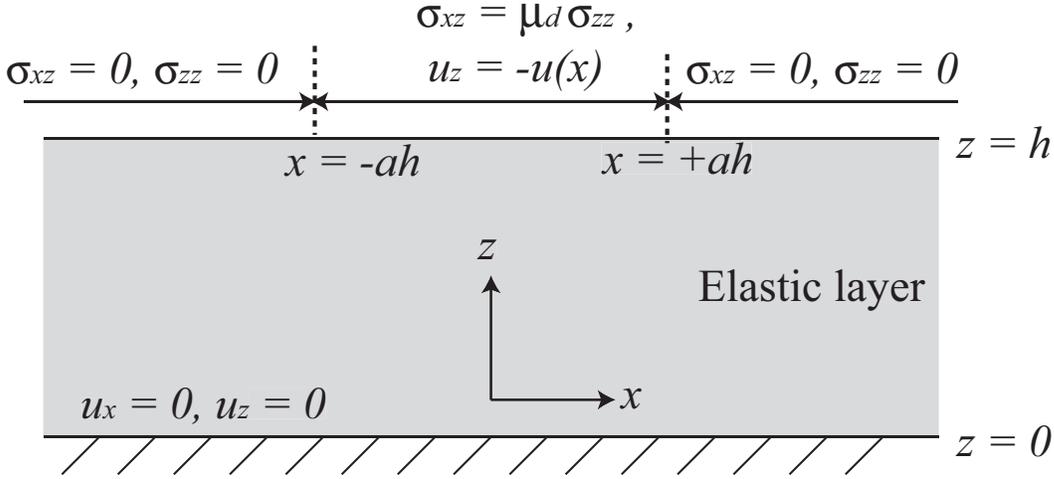}
\caption{Sketch of the system considered in the exact model. An elastic film is perfectly adhering on its rigid base ($z=0$). At its surface ($z=h$) it is stress free outside of the contact region ($\left|x\right|<ah$), with $a$ being a result of the calculation. Within the contact region, the normal displacements $u_z$ are prescribed and in steady sliding $\sigma_{xz}=\mu_d \sigma_{zz}$ is assumed everywhere at the interface, $\mu_d$ being the dynamic friction coefficient.\label{schemaKA}}
\end{figure}

To allow for a direct quantitative comparison with the previous experimental stress profiles we have developed the following bidimensional exact model (Fig. \ref{schemaKA}). A linear incompressible elastic film, of thickness $h$ and Young's modulus $E$, is loaded under plane strain conditions by a rigid circular body of radius $R$ moving at a constant velocity $V$. We postulate quasi-static motion, i.e. the characteristic time $h/c$ for sound waves of velocity $c$ to travel across the film is assumed to be smaller than the characteristic time $a/V$ associated with the indentor motion, so that the elastic film is at equilibrium at all times. The problem is made dimensionless by expressing the coordinates ($x,z$), displacements $u_i(x,z)$ and stress $\sigma_{ij}(x,z)$ in units of $h$, ${h^2}/{2R}$ and ${Eh}/{6R}$, respectively.

The constitutive equations for the elastic film can be written as
\begin{equation}
\sigma_{ij}=-\Sigma\delta_{ij}+\frac{\partial u_i}{\partial x_j}+ \frac{\partial u_j}{\partial x_i} \label{eq-elasticity}
\end{equation}
where $\Sigma$ is the pressure. The equilibrium equations in the film and the condition of incompressibility are
\begin{align}
\nabla \Sigma &= \triangle \vec{u} \label{eq-poisson}\\
\nabla\cdot\vec{u} &= 0 \label{eq:bulk2}
\end{align}
We specify the following boundary conditions
\begin{align}
                  u_x(x,0) = u_z(x,0)&=0     & \label{eq-bc-base}\\
\sigma_{xz}(x,1)+\mu_d \sigma_{zz}(x,1)&=0   & \label{eq-bc-coulomb}\\
                     \sigma_{zz}(x,1)&=0     & \mbox{for}\;\; |x|>a \label{eq-bc-freesurf}\\
                             u_z(x,1)&=-u(x) & \mbox{for}\;\; |x|<a \label{eq-bc-contact}
\end{align}
where $z=0$ and $z=1$ correspond to the locations of the base and the surface of the elastic film respectively. Eq. (\ref{eq-bc-base}) accounts for the perfect adhesion of the film to its rigid base, Eq. (\ref{eq-bc-coulomb}) corresponds to Amontons' law of friction with a dynamic friction coefficient $\mu_d$, Eq. (\ref{eq-bc-freesurf}) insures that the surface of the film is traction-free outside the contact zone and Eq. (\ref{eq-bc-contact}) defines the normal displacement induced by the indentor over the contact zone of width $2a$. For a circular rigid indentor the normal displacement has a parabolic profile given by 
\begin{equation}
u(x)=\frac{1}{\alpha} - (x-x_0)^2
\end{equation}
where $x_0$ represents the asymmetry of the steady sliding contact and $\alpha=h^2/2R \delta$ with $\delta$ being the normal displacement of the indentor. Both $x_0$ and $\alpha$ are selected by the system for a given width of the contact zone area $a$ and friction coefficient $\mu_d$.

As suggested by the strip geometry and the boundary conditions, the resolution involves the use of Fourier sine and cosine transforms (\cite{AddaBedia-BenAmar-PRL-2001}). Any spatial distribution function $D(x,y)$ of the problem (displacement, strain or stress) may be decomposed into
\begin{equation}
D(x,y)=\int_{0}^{\infty} D^{(c)}(k,y)\cos kx dk+ \int_{0}^{\infty} D^{(s)}(k,y)\sin kx dk
\end{equation}
Substituting this representation into the bulk equations Eqs. (\ref{eq-elasticity})--(\ref{eq:bulk2}) and the boundary conditions (\ref{eq-bc-base})--(\ref{eq-bc-contact}) and exploiting the parity properties of the sine and cosine functions, lead to the following equations
\begin{align}
\int_{0}^{\infty} \sigma_{zz}^{(c)}(k,1)\cos kx \,dk&=0\qquad |x|>a 
\label{eq:bc1}\\
\int_{0}^{\infty} \sigma_{zz}^{(s)}(k,1)\sin kx \,dk&=0\qquad |x|>a
\label{eq:bc2}
\end{align}
and
\begin{align}
&&\int_{0}^{\infty} \left[F_0(k)\sigma_{zz}^{(c)}(k,1)+\mu_d F_1(k)\sigma_{zz}^{(s)}(k,1)\right] \frac{\cos kx}{2k} \,dk \nonumber\\
&&=-\frac{1}{2}\left[u(x)+u(-x)\right]\qquad |x|<a \label{bc3}\\
&&\int_{0}^{\infty} \left[-\mu_d F_1(k)\sigma_{zz}^{(c)}(k,1)+F_0(k)\sigma_{zz}^{(s)}(k,1)\right] \frac{\sin kx}{2k} \,dk \nonumber\\
&& =-\frac{1}{2}\left[u(x)-u(-x)\right]\qquad |x|<a \label{bc4}
\end{align}
where
\begin{align}
F_0(k)&=\frac{\sinh(2k)-2k}{\cosh(2k)+1+2k^2}, \label{f0}\\
F_1(k)&=\frac{2 k^2}{\cosh(2k)+1+2k^2}.\label{f1}
\end{align}

The conditions (\ref{eq:bc1}, \ref{eq:bc2}) are identically satisfied by
\begin{align}
\sigma_{zz}^{(c)}(k,1) &= \int_{0}^{a} \phi(t)J_0(k t) dt \label{phi} \\
\sigma_{zz}^{(s)}(k,1) &= \int_{0}^{a} t\psi(t)J_1(k t) dt \label{psi}
\end{align}
irrespective of $\phi(t)$ and $\psi(t)$, with $J_0(x)$ and $J_1(x)$ being the Bessel functions of the first kind. The functions $\phi(t)$ and $\psi(t)$ now become the unknowns in the problem.

In two-dimensional contact problems, the indentation depth is undeterminate, which requires differentiating the boundary conditions (\ref{bc3})--(\ref{bc4}) with respect to $x$ before replacement into the representation (\ref{phi})--(\ref{psi}). One then classically gets a set of coupled integral
equations (see e.g. \cite{Spence-JElasticity-1975} and \cite{Gladwell-Springer-1980}), that are here of Abel type which fix the functions $\phi(t)$ and $\psi(t)$. Inverting this set of equations using the Abel transform yields
\begin{align}
\phi(x)+\int_{0}^{a} M_{00}(x,t)\phi(t)\,dt + \mu_d \int_{0}^{a}M_{10}(x,t)\psi(t)dt&=-4x \label{ab1} \\
\psi(x)+\mu_d \int_{0}^{a} M_{01}(x,t)\phi(t)dt - \int_{0}^{a} M _{11}(x,t)\psi(t)dt&=0 \label{ab2}
\end{align}
where $M_{ij}(x,t)=(-1)^j x^{1-j} t^i \int_0^\infty k (F_{\left|i-j\right|}(k)-\delta_{ij})J_i(kt)J_j(kx)dk$. Eqs. (\ref{ab1})-(\ref{ab2}) are independent of the parameters $x_0$ and $\alpha$ which allows to solve them once the constant $\mu_d$ and $a$ are fixed. This simplifies the numerical scheme.
Then, $x_0$ and $\alpha$ are fixed \textit{a posteriori} by using Eq. \ref{bc3} and the derivative of Eq. \ref{bc4} with respect to $x$ at, say $x=0$. This leads to the following equations
\begin{align}
4x_0=& \mu_d\int_0^a \phi(t)\int_0^\infty F_1(k)J_0(kt)dk\,dt -\int_0^a t\psi(t)\int_0^\infty F_0(k)J_1(kt)dk\,dt\\
\frac{1}{\alpha}=& x_0^2 - \int_0^a \phi(t)\int_0^\infty \frac{F_0(k)}{2k}J_0(kt)dk\,dt -\mu_d\int_0^a t\psi(t)\int_0^\infty \frac{F_1(k)}{2k}J_1(kt)dk\,dt
\end{align}

The displacement and stress fields can be easily expressed as functions of $\phi(x)$, $\psi(x)$, $x_0$ and $\alpha$ and thus can also be calculated numerically. The
lineic normal load $P_L$ applied to the film surface can then be calculated using the following expression
\begin{equation}
P_L=-\int_{-a}^{a} \sigma_{zz}(x,1)\,dx=-\pi \int\limits_0^a
{\phi(t)dt}. \label{force}
\end{equation}
Using the constitutive equations and providing simple algebraic transformations the normal stress $\sigma_{zz}(x,0)$ and the tangential stress $\sigma_{xz}(x,0)$ at the rigid base are given by
\begin{align}
 \sigma_{zz}(x,0) &= \int\limits_0^a {\left[ {Z_1(x,t) + \mu_d Z_3(x,t)} \right] \phi(t)dt} - \int\limits_0^a {\left[ {  \mu_d Z_2(x,t) - Z_4(x,t)} \right] t \psi(t)dt}  \\
 \sigma_{xz}(x,0) &= -\int\limits_0^a {\left[ {\mu_d Z_5(x,t) + Z_3(x,t)} \right] \phi(t)dt} + \int\limits_0^a {\left[ {Z_2(x,t) - \mu_d Z_6(x,t)} \right]t \psi(t)dt}
\end{align}
where the kernels $Z_i(x,t)$ are explicitly
\begin{align}
Z_1(x,t) &= \int\limits_0^\infty  {A( k)\cos( {kx})J_0( {kt})dk},\label{z1}\\
Z_2(x,t) &= \int\limits_0^\infty  {B( k)\cos( {kx})J_1( {kt})dk},\label{z2}\\
Z_3(x,t) &= \int\limits_0^\infty  {B( k)\sin( {kx})J_0( {kt})dk},\label{z3}\\
Z_4(x,t) &= \int\limits_0^\infty  {A( k)\sin( {kx})J_1( {kt})dk},\label{z4}\\
Z_5(x,t) &= \int\limits_0^\infty  {C( k)\cos( {kx})J_0( {kt})dk},\label{z5}\\
Z_6(x,t) &= \int\limits_0^\infty  {C( k)\sin( {kx})J_1( {kt})dk},\label{z6}
\end{align}
with $A(k)$, $B(k)$ and $C(k)$ being
\begin{align}
A(k)&= \frac{2(\cosh(k)+k\sinh(k))}{\cosh(2k)+1+2k^2},\label{ak}\\
B(k)&= \frac{2k\cosh(k)}{\cosh(2k)+1+2k^2},\label{bk}\\
C(k)&= \frac{2(\cosh(k)-k\sinh(k))}{\cosh(2k)+1+2k^2}.\label{ck}
\end{align}

In practice, the input parameters of the model are chosen to be $\mu_d$ and $P_L$, and the resulting normal and tangential stress profiles at the base of the film are derived.

\section{Discussion}\label{discussion}

We recall here that the calculation presented in the previous section is the first one relaxing Goodman's assumption for the frictional steady sliding of a layered material. In order to assess the impact of this increment on the mechanical description of such contacts, we directly compare, for various combinations of the input parameters $\mu_d$ and $P_L$, the stress profiles obtained from both our exact calculation and an additional calculation derived along the same lines as the exact one but with Goodman's assumption. The latter test model, referred to as Goodman's model is detailed in Appendix \ref{appendix}.

\begin{figure}[ht!]
\includegraphics[width=\columnwidth]{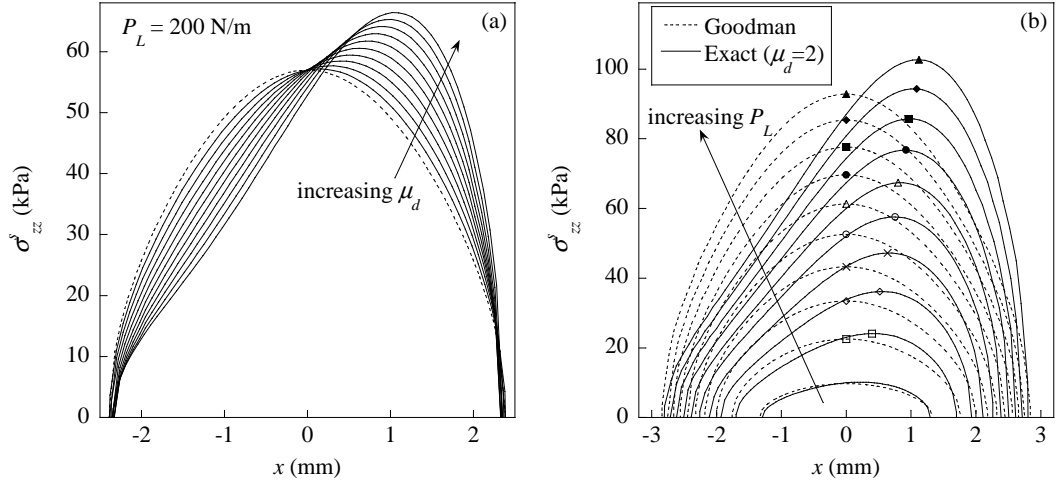}
\caption{Normal stress profiles $\sigma^s_{zz}=\sigma_{zz}(x,1)$ at the surface of the film, calculated with the exact model (solid lines) or with Goodman's model (dashed lines). (a) $\mu_d$ increases from 0.3 to 3.0 with steps of 0.3 for the same lineic normal load $P_L$\,=\,200\,N.m$^{-1}$. For all cases, the contact radius is 2.36\,$\pm$\,0.03\,mm. (b) $P_L$ increases from 20 to 380\,Pa.m$^{-1}$ with steps of 40\,Pa.m$^{-1}$ for the same friction coefficient $\mu_d$\,=\,2.0. For the exact model, contact widths are 2.60, 3.40, 3.88, 4.24, 4.52, 4.80, 5.02, 5.24, 5.44 and 5.62\,mm respectively. For Goodman's model, contact widths are 2.68, 3.52, 4.00, 4.36, 4.66, 4.90, 5.12, 5.32, 5.52 and 5.68\,mm respectively. For all these graphs, the following parameters were used: $E$\,=\,2.2\,MPa, $R$\,=\,130\,mm, $h$\,=\,2\,mm.\label{surface}}
\end{figure}

Figure \ref{surface} shows the normal stress profiles $\sigma^s_{zz}=\sigma_{zz}(x,1)$ at the surface of the film. For each normal stress profile, the corresponding tangential stress is obtained by multiplying the former by the friction coefficient $\mu_d$, i.e. $\sigma^s_{xz}=\mu_d \sigma^s_{zz}$ following Amontons' law - see Eq. \ref{eq-bc-coulomb}. As expected, for $\mu_d=0$, the exact calculation matches Goodman's result and yields symmetric fields with an integral (area below the curve) equal to $P_L$. For increasing $\mu_d$ at constant $P_L$, the profiles maintain their integral while becoming increasingly asymmetric, with a growing maximum shifting towards the leading edge of the moving indentor. A similar behavior for the envelope is observed for an increasing $P_L$ at constant $\mu_d$. Interestingly, Goodman's model deviates significantly from the exact one, even in the favourable situation considered here where the material is incompressible and the film is relatively thick.

\begin{figure}[ht!]
\includegraphics[width=\columnwidth]{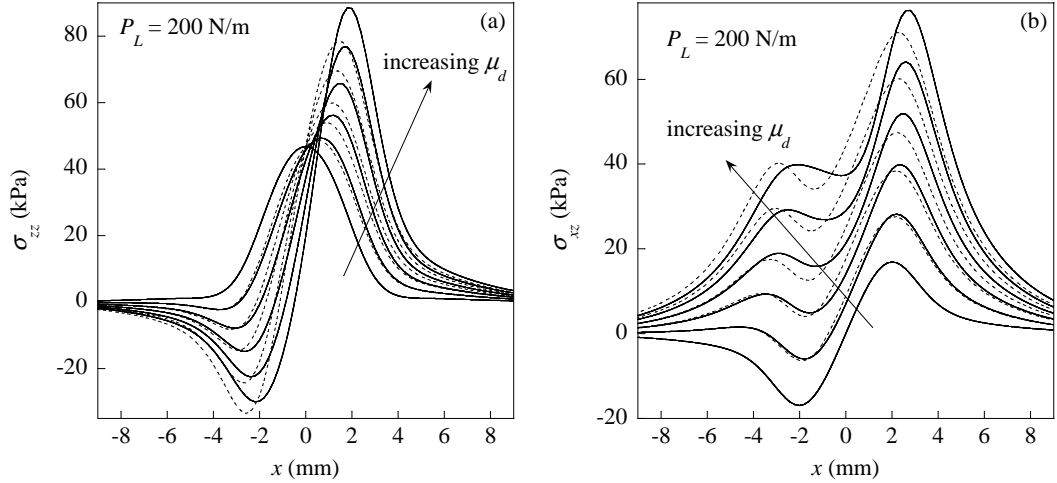}
\caption{(a) Normal stress profiles $\sigma_{zz}(x,0)$ and (b) tangential stress profiles along the direction of movement $\sigma_{xz}(x,0)$ calculated at the base of the elastic film with the exact model (solid lines) or with Goodman's model (dashed lines). $\mu_d$ increases from 0 to 3.0 with steps of 0.6 for the same lineic normal load $P_L$\,=\,200\,N.m$^{-1}$. The contacts widths are equal to that given in the legend of Fig. \ref{surface}(a). The following parameters were used: $E$\,=\,2.2\,MPa, $R$\,=\,130\,mm, $h$\,=\,2\,mm.\label{bottom1}}
\end{figure}

\begin{figure}[ht!]
\includegraphics[width=\columnwidth]{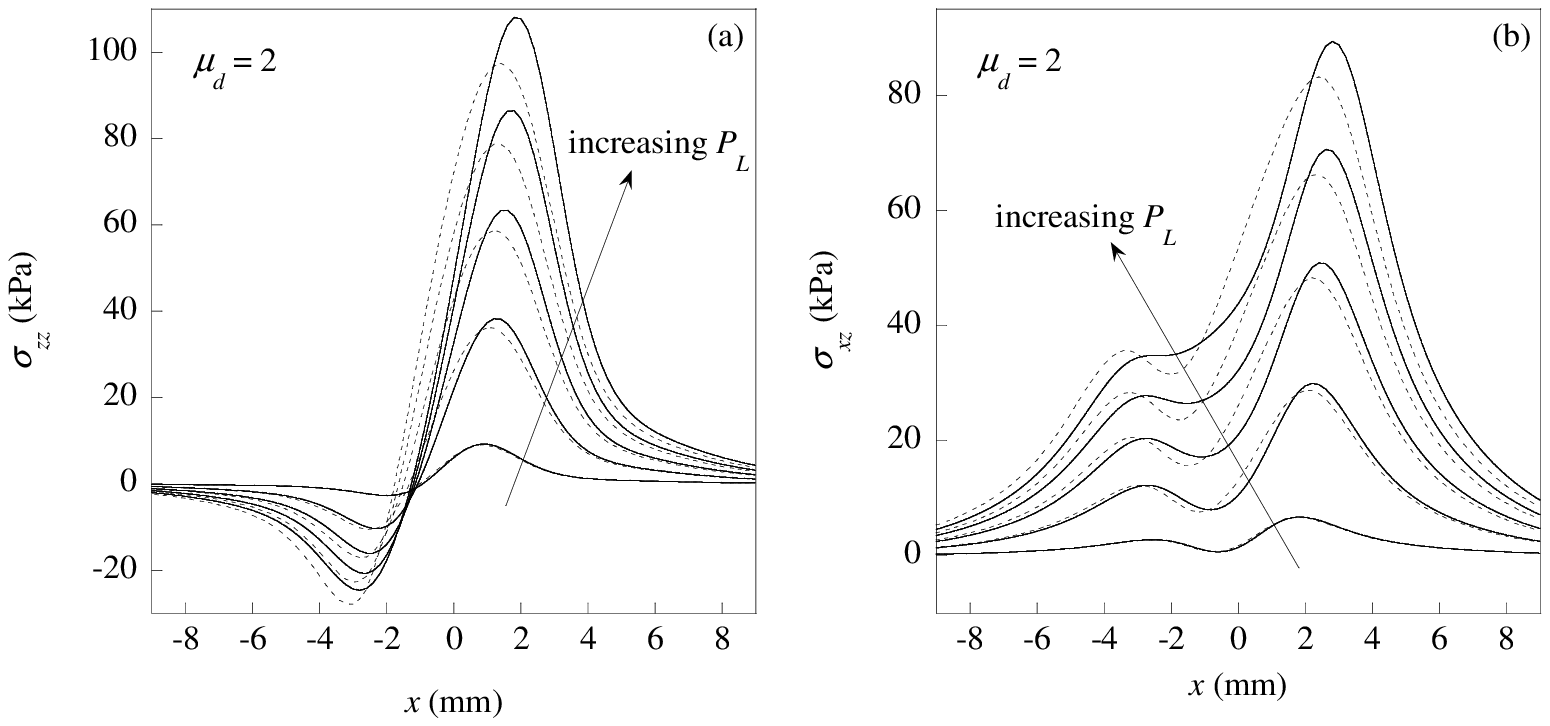}
\caption{(a) Normal stress profiles $\sigma_{zz}(x,0)$ and (b) tangential stress profiles along the direction of movement $\sigma_{xz}(x,0)$ calculated at the base of the elastic film with the exact model (solid lines) or with Goodman's model (dashed lines). $P_L$ increases from 20 to 380\,Pa.m$^{-1}$ with steps of 80\,Pa.m$^{-1}$ for the same friction coefficient $\mu_d$\,=\,2.0. The contact widths are equal to that given in the legend of Fig. \ref{surface}(b). The following parameters were used: $E$\,=\,2.2\,MPa, $R$\,=\,130\,mm, $h$\,=\,2\,mm.\label{bottom2}}
\end{figure}

Figure \ref{bottom1} shows both the normal and tangential stress profiles, $\sigma_{zz}(x,0)$ and $\sigma_{xz}(x,0)$, at the base of the film, where the the stress $\sigma$ is actually measured. $\sigma$ is related to $\sigma^s$ at the free surface of the film through a convolution with the Green function for an elastic membrane of thickness $h$. Since the latter has a typical width $\simeq h$, $\sigma$ cannot exhibit spatial modulations over length scales smaller than $h$=\,2\,mm. The spatial resolution of the MEMS ($\simeq$\,1\,mm) is therefore sufficient to probe the stress field $\sigma$ at the base of the elastic film. For $\mu_d$\,=\,0, the normal stress profile is symmetric with an integral equal to $P_L$ whereas the tangential stress profile is antisymmetric with a vanishing integral. For a given lineic load $P_L$, an increasing $\mu_d$ qualitatively results in growing additional contributions to the profiles, anti-symmetric for the normal stress and symmetric for the tangential stress. The integral of the normal stress profile remains equal to $P_L$ while the integral of the tangential stress profile becomes $\mu_d P_L$. Similar features are observed on Fig. \ref{bottom2}, which shows $\sigma_{zz}(x,0)$ and $\sigma_{xz}(x,0)$ for an increasing lineic load $P_L$ and a given friction coefficient $\mu_d$. Goodman's model yields qualitatively similar results but with growing errors for increasing $P_L$ or $\mu_d$. In particular, Goodman's model underestimates the amplitude of the maxima of both stress components at positive $x$ and overestimates the amplitude of both the negative part of the normal stress and the dip in the tangential profiles at negative $x$.

The measured stress profiles $S_{zz}(x)$ and $S_{xz}(x)$ along $y$\,=\,0 can be now quantitatively compared to the stress profiles  $\sigma_{zz}(x)$ and $\sigma_{xz}(x)$ calculated at the base of the elastic film and convoluted with the apparatus functions $G_{zz}$ and $G_{xz}$ determined in Section \ref{Set-up}. In the limit of a bidimensional geometry, the input parameters used in the calculation - namely the applied lineic load $P_L$ and the dynamic friction coefficient $\mu_d$ - should ideally be deduced from the macroscopic measurements of $P$ (the normal load) and $Q$ (the tangential load) by using $P/L$ and $Q/P$ respectively, with $L$ being the contact length. This approach yields inconsistent stress profiles for two reasons. First, with our finite sized punch experimental system, the contribution of edge effects to the total normal load $P$ is not negligible. For a given $x$, the interfacial pressure has a minimum around $y$\,=\,0, so that $P/L$ over-estimates the effective lineic load at the location of the measured profile. Second, the measured macroscopic friction coefficient $Q/P$ turns out to be a decreasing fonction of $P$ (and thus of the local pressure), assuming values from 1.5$\pm$0.1 at $P$\,=\,0.34\,N down to 1.36$\pm$0.04 at $P$\,=\,2.75\,N, which are typical for PDMS on glass steady sliding contacts (see e.g. \cite{Galliano-Bistac-Schultz-JColloidInterfaceSci-2003, Wu-Clain-Buguin-DeGennes-Brochard-JAdhesion-2007}). These averaged values under-estimate the effective friction coefficient at the location of the measured profile since the pressure has a minimum around $y$\,=\,0. To circumvent this difficulty, we extracted $P_L$ and $\mu_d$ from the measured stress profiles as $P_L=\int^{\infty}_{-\infty}S_{zz}dx$ and $\mu_d=\int^{\infty}_{-\infty} S_{xz}dx / \int^{\infty}_{-\infty} S_{zz}dx$. With such definitions, $P_L$ is found to increase from $20$ to $220 \, N.m^{-1}$ and $\mu_d$ to decrease from $2.6$ to $2.0$ when $P$ varies from $0.34$ to $2.75 \, N$.

\begin{figure}[ht!]
\includegraphics[width=\columnwidth]{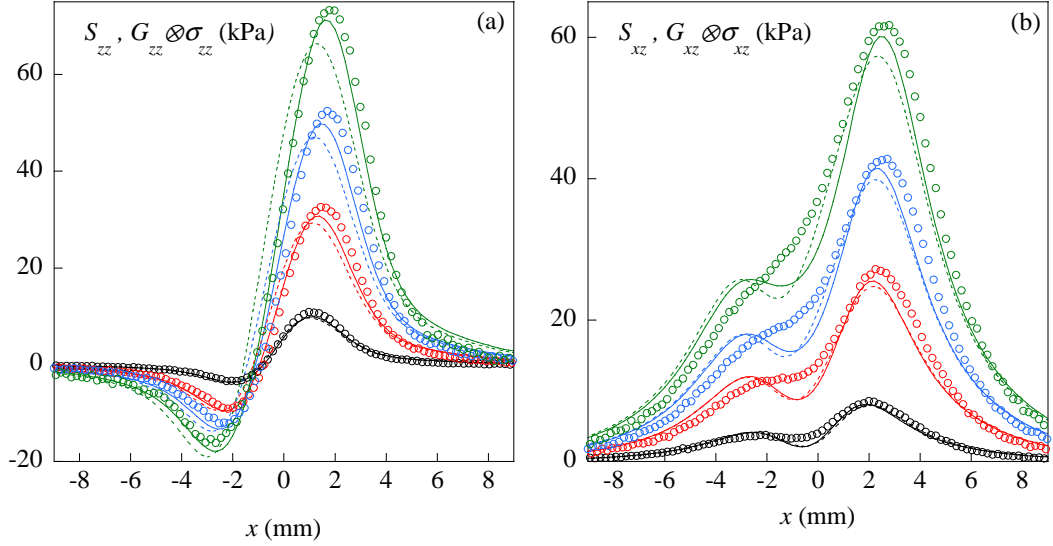}
\caption{Measured stress profiles ($\circ$, for clarity only one percent of the data points is shown) at $y$\,=\,0 (a) $S_{zz}$ and (b) $S_{xz}$ in steady sliding regime for increasing normal loads ($P$\,=\,0.34\,N in black, 1.03\,N in red, 1.72\,N in blue and 2.75\,N in green) and $V$\,=\,1\,mm.s$^{-1}$. Comparison is made with (a) $G_{zz} \otimes \sigma_{zz}$ and (b) $G_{xz} \otimes \sigma_{xz}$ where $\sigma_{zz}$ and $\sigma_{xz}$ are computed from the exact model (solid lines) or from Goodman's model (dashed lines).\label{compa}}
\end{figure}

Figure \ref{compa} shows the measured profiles together with the predicted stress profiles convoluted with the apparatus functions, for both our exact model and Goodman's model. The two calculations predict profiles in reasonable agreement with the experimental ones. In particular, they account for both the negative part of $S_{zz}(x)$ and the dip of $S_{xz}(x)$ at negative $x$. In order to quantify the deviations between the experimental and calculated profiles, we compute the quantity $\chi = \sqrt{\Sigma_i (E_i-C_i)^2 / \Sigma_i E_i^2}$, where $E_i$ are the experimental data points and $C_i$ are the calculated ones. For the tangential stress, both models yield similar values of $\chi$=11\,$\pm$\,1$\%$, with no clear load dependance. For the normal stress profiles, the exact model yields an almost constant $\chi$=11\,$\pm$\,3\,$\%$ over the range of normal loads $P$ explored. For Goodman's model, $\chi$ increases with the load, between 12 and 28\,$\%$, indicating a decreasingly good fit to the experimental data with increasing $P$. The exact model is therefore the one that follows most closely the evolution of the experimental profiles with increasing normal load (Fig. \ref{compa}), which is consistent with the fact that Goodman's assumption is expected to fail as the ratio of contact size $a$ to film thickness $h$ becomes large.

Although the exact model accounts for the data better, non-negligible robust deviations are observed for which we do not have any definitive explanation. Two central assumptions used in both models are however amenable to refinement and may explain the observed deviations. First, the interface is assumed to be molecularly smooth whereas the surface of the elastomer exhibits a micrometric roughness. The resulting multicontact interface is thus expected to exhibit finite compressive and shear compliances. This feature has been shown to modify, with respect to smooth contacts, both the stress (\cite{Greenwood-Tripp-JApplMech-1967, Scheibert-Prevost-Frelat-Rey-Debregeas-EPL-2008, Chateauminois-Fretigny-EurPhysJE-2008}) and displacement (\cite{Scheibert-Debregeas-Prevost-CondMat-2008}) fields. These effects are expected to induce vanishing corrections at increasingly high loads. The second questionable assumption is the existence of a single pressure-independent friction coefficient. This is clearly at odds with the observed decrease of $Q / P$ as a function of $P$. Such a behavior is usually attributed to the finite adhesion energy of the interface (e.g. \cite{Carbone-Mangialardi-JMechPhysSolids-2004}), and is sensitive to the geometrical properties of the film roughness.

\section{Conclusion}\label{conclusion}

This work provides the first spatially resolved direct measurement of the stress field at a sliding contact. The choice of a cylinder-on-plane geometry has allowed us to quantitatively compare the profiles measured at the center line of the contact with bidimensional calculations. An exact model was developed to predict the stress field at the sliding contact assuming linear elasticity and a locally valid Amontons' friction law, but without the classical Goodman's assumption on the normal displacements. This model correctly captures the measured stress profiles with typical deviations of less than 14\,$\%$. In the range of loads explored experimentally, this calculation does not differ drastically from the classical calculation involving Goodman's assumption. However, the present model is expected to provide significant improvements over Goodman's model as the thickness of the film is further reduced or as the load is further increased. In these cases, Goodman's assumption becomes increasingly inaccurate.

Robust deviations between the experiments and the model have been briefly discussed along two lines, namely the finite compliance of the multicontact interface and the pressure-dependence of the friction coefficient. However, the cylinder-on-plane experiment described here, which was specifically designed to allow for a comparison with bidimensional models, is not best suited to study such fine effects. As discussed, the resulting edge effects do not allow one to use well-controlled or measured macroscopic quantities, e.g. $P$ and $Q$, as input parameters in the models. This could be done for instance with a sphere-on-plane geometry, but it would require for comparison a more complex 3D stress analysis. Work in this direction is in progress.

\bibliographystyle{elsart-harv}

\appendix
\section{Goodman's model}
\label{appendix}

The calculation scheme involves first solving the exact model decribed in section \ref{model}, but with $\mu_d=0$, to obtain the corresponding interfacial (symmetric) pressure field $p_0(x)$. The second step is to solve the same constitutive equations for the following new boundary conditions:
\begin{align}
                    u_x(x,0) = u_z(x,0)&=0      & \label{eq:bc0}\\
\sigma_{xz}(x,1)+\mu_d \sigma_{zz}(x,1)&=0      & \label{eq:bc3}\\
                       \sigma_{zz}(x,1)&=0      & \mbox{for}\;\; |x|>a \label{eq:bcsigma1}\\
                       \sigma_{zz}(x,1)&= -p_0(x) & \mbox{for}\;\; |x|<a \label{eq:bcsigma2}
\end{align}
where $z=0$ and $z=1$ correspond to the locations of the base and the surface of the elastic film respectively and $p_0(x)$ is the pressure field that results from the first step. Eq. (\ref{eq:bc0}) accounts for the perfect adhesion of the film to its rigid base, Eq. (\ref{eq:bc1}) corresponds to Amontons' law of friction with a dynamic friction coefficient $\mu_d$ and Eq. (\ref{eq:bcsigma1}) insures that the surface of the film is traction-free outside the contact zone of width $2a$. Eq. (\ref{eq:bcsigma2}) corresponds to Goodman's assumption which implies that the interfacial normal stress field is not affected by frictional stress, and so $p_0(x)$ from the previous step is used.

The Fourier transform of Eqs. (\ref{eq:bcsigma1})--(\ref{eq:bcsigma2}) yields
\begin{align}
 \int_{0}^{\infty} \sigma_{zz}^{(c)}(k)\cos kx \,dk&=0     & \mbox{for}\;\; |x|>a\;, \label{eq:a1}\\
\int_{0}^{\infty} \sigma_{zz}^{(c)}(k) \cos kx \,dk&=-p(x) & \mbox{for}\;\; |x|<a\;.\label{eq:a2}
\end{align}
Eq. (\ref{eq:a1}) is identically satisfied by
\begin{equation}
\sigma_{zz}^{(c)}(k)=\int_{0}^{a} F(t)J_0(k t) dt\; \label{bessel}
\end{equation}
where $J_0(x)$ is the Bessel function of the first kind. By replacing (\ref{bessel}) into Eq. (\ref{eq:a2}) we get the following integral equations that determine the function $F(t)$
\begin{equation}
\int_{x}^{a}\frac{F(t)}{\sqrt{t^2-x^2}}\,dt =-p(x) \equiv \int_{x}^{a}\frac{\phi(t)}{\sqrt{t^2-x^2}}\,dt \qquad |x|<a\;, 
\label{eq:phi}
\end{equation}
where $\phi$ is the function defined in Eq. (\ref{phi}) (see section \ref{model}) obtained for the particular case where $\mu_d=0$. The solution for $F(x)$ is readily given by $F(x)=\phi(x)$.

The normal stress $\sigma_{zz}(x,0)$ and the tangential stress $\sigma_{xz}(x,0)$ at the rigid base are then given by 
\begin{align}
\sigma_{zz}(x,0) &=  \int\limits_0^a \left[ {Z_1(x,t) + \mu_d Z_3(x,t)} \right] \phi(t)dt\;, \\
\sigma_{xz}(x,0) &=-  \int\limits_0^a \left[ {\mu_d Z_5(x,t) + Z_3(x,t)} \right]\phi(t)dt\;,
\end{align}
where the kernels $Z_i(x,t)$ and $A(k)$, $B(k$ and $C(k)$ are given by Eqs. (\ref{z1})--(\ref{ck}).
\end{document}